\documentclass[12pt]{article}
\usepackage{epsf}
\usepackage{psfig,epsfig}
\def\Slash #1{\hbox{$#1$}\mkern-11.0mu\lower 0.pt \hbox{/}} 
\def\Lc #1{\cal L} 
\def\bea{\begin{eqnarray}}
\def\eea{\end{eqnarray}}
\def\beq{\begin{equation}}
\def\eeq{\end{equation}}
\def\ba{\begin{array}}
\def\ne{\hbox{$\nu_e$ }}
\def\nm{\hbox{$\nu_\mu$ }}
\def\nt{\hbox{$\nu_\tau$ }}
\def\mnt{\hbox{$m_{\nu_\tau}$ }}
\def\eq#1{{eq. (\ref{#1})}}

\def\lsim{\raise0.3ex\hbox{$\;<$\kern-0.75em\raise-1.1ex\hbox{$\sim\;$}}}
\newcommand {\ignore}[1]{}
\evensidemargin=\oddsidemargin
\begin{document}
\begin{titlepage}
\samepage{
\rightline{\tt hep-ph/9906343}
\rightline{\tt IFIC/99-35}
\rightline{\tt FTUV/99-34}
\rightline{\tt FSU-HEP-990529}
\rightline{\tt June 1999}
\vfill 
\begin{center}
   {\Large \bf Unification of gauge couplings and the tau neutrino\\
   \vskip .2cm
   mass in Supergravity without R--parity}
\vfill
\vspace{.08in}
   {\large
     M. A. D\'\i az   $^1$,$\,$
     J. Ferrandis $^2$,$\,$
     J. C. Rom\~ao  $^3$,$\,$
     J. W. F. Valle $^2$}
\vspace{.18in}

{\it $^1$ High Energy Group, Physics Department, Florida State University\\ 
          Tallahassee, Florida 32306, USA\\}
{\it $^2$ Departament de F\'\i sica Te\`orica, IFIC--CSIC,
          Universitat de Val\`encia\\
          46100 Burjassot, Val\`encia, Spain \\ }
{\it $^3$ Departamento de F\'\i sica, Instituto Superior T\'ecnico\\
          Av. Rovisco Pais, 1, $\:\:$ 1049-001 Lisboa, Portugal \\}

\end{center}

\begin{abstract}
  
{\rm Minimal R--parity violating supergravity predicts a value for
$\alpha_s(\hbox{M}_Z)$ smaller than in the case with conserved the
R--parity, and therefore closer to the experimental world average.  We
show that the R--parity violating effect on the $\alpha_s$ prediction
comes from the larger two-loop b-quark Yukawa contribution to the
renormalization group evolution of the gauge couplings which
characterizes R--parity violating supergravity. The effect is
correlated to the tau neutrino mass and is sensitive to the initial
conditions on the soft supersymmetry breaking parameters at the
unification scale. We show how a few percent effect on 
$\alpha_s(\hbox{M}_Z)$ may naturally occur even with \nt masses as small 
as indicated by the simplest neutrino oscillation interpretation of the 
atmospheric neutrino data from Super-Kamiokande.}
\end{abstract}
\vspace{.13in}  
}
\vfill
\noindent 
email addresses:\\ diaz@hep.fsu.edu,\\
 ferrandi@flamenco.ific.uv.es,\\
 fromao@alfa.ist.utl.pt, \\
 valle@flamenco.ific.uv.es\\

\end{titlepage}
\setcounter{footnote}{0}
\baselineskip=24pt

\section{Introduction}

The prediction for the strong gauge coupling constant $\alpha_s
(\hbox{M}_Z)$ is one of the milestones of unification models
\cite{alphabeyond}. Recent studies of gauge coupling unification 
in the context of minimal R-parity conserving supergravity
\cite{Langacker:1995fk,Langacker:1993rq,Carena:1993ag} agree that
using the experimental values for the electro-magnetic coupling and
the weak mixing angle the prediction obtained for $\alpha_s
(\hbox{M}_Z) \approx 0.129 \pm 0.010$ \cite{Langacker:1995fk} is about
2$\sigma$ larger than indicated by the most recent world average value
$\alpha_s(\hbox{M}_Z)^{W.A.}=0.1189\pm 0.0015$ \cite{Caso:1998tx}.

Here we re-consider the $\alpha_s$ prediction in supergravity (SUGRA).
In addition to the standard MSUGRA we consider simplest supergravity
version with a bi-linear breaking of R parity
\cite{epsrp:1997,BRpVmas,NillHempJoshi,Bisset:1998bt,Nowak:1996dx}.  
This model is theoretically motivated by the fact that it provides
parametrization of many of the features of a class of models in which
R--parity breaks spontaneously due to a sneutrino vacuum expectation
value (vev) ~\cite{rpos}. Moreover, in the simplest case where
R--Parity violation lies only in the third generation, the model
coincides with the most general explicit R--parity violating model and
provides its simplest description.

One of the main features of R--parity violating models is the
appearance of masses for the neutrinos~\cite{rpos,rpre}. As a result,
these models have attracted a lot of attention \cite{rprecent,rpfull}
since the latest round of Super-Kamiokande results \cite{Fukuda}.  

In this paper we show that in the simplest SUGRA R--parity breaking
model, with the same particle content as the MSSM and with no new
interactions (such as trilinear R--parity breaking couplings), there
appears an additional negative contribution to $\alpha_s$, which can
bring the theoretical prediction closer to the experimental world
average. This additional contribution to $\alpha_s$ comes from
two--loop b-quark Yukawa effects on the renormalization group equation
(RGE) for $\alpha_s$.  Moreover, we show that this contribution is
typically correlated to the tau--neutrino mass which is induced by
R--parity breaking and which controls the R--parity violating
effects. We also discuss this correlation within different models for
the initial conditions on the soft supersymmetry breaking parameters
at the unification scale.  We show how to obtain a sizeable effect on
$\alpha_s(\hbox{M}_Z)$ even with \nt masses as small as indicated by
the simplest neutrino oscillation interpretation of the atmospheric
neutrino data from Super-Kamiokande.

\section{The MSSM Renormalization Group Equations}

The two loop renormalization group equations \cite{2loopRGE} for the
gauge coupling constants in the MSSM have the form
\begin{equation} 
\frac{dg_i}{dt}=\frac{g_i}{16\pi ^2}\left( b_i g_i^2+\frac 1{16\pi 
^2}\left( \sum_{j=1}^3b_{ij}\,g_i^2g_j^2-\sum_{l=t,b,{\tau }%
}b_{il}^{\prime}\,g_i^2h_l^2\right) \right)  
\label{giRGE}
\end{equation}
where $g_i$, $i=1,2,3$, are the gauge couplings of the $U(1)$,
$SU(2)$, and $SU(3)$ groups respectively, and $h_l$, $l=t,b,\tau$, are
the quark and lepton Yukawa couplings of the third generation. The
numerical coefficients $b_i$, $b_{ij}$, and $b'_{il}$ are given in
ref.~\cite{2loopRGE}.

It is useful to obtain an approximate analytical solution to the gauge 
coupling constants from eq.~(\ref{giRGE}). This is done by neglecting 
the two loop Yukawa contribution in first approximation. The result is 
\cite{Hall:1980}
\begin{equation}
\frac{1}{\alpha_i(\mu)}=
\frac{1}{\alpha_U(\hbox{M}_U)}+b_i t+
\frac{1}{4\pi}\sum_{j=1,2,3}\frac{b_{ij}}{b_i}
\ln \left[1+b_j \alpha_U(\hbox{M}_U)t \right]
-\Delta_i
\label{alfaiApp}
\end{equation}
where $t=\frac{1}{2\pi}\ln \left(\hbox{M}_U/\mu\right)$, $\alpha_U$ is
the unified gauge coupling constant, $M_U$ is the unification scale,
$\mu$ is an arbitrary scale, and $\Delta_i$ are corrections due to
several effects, mainly threshold corrections. Although GUT-type
threshold corrections are potentially sizeable, we neglect them here
since they are in general model-dependent. For a discussion see ref.
\cite{Langacker:1995fk,Langacker:1993rq,Bagger:1995bw}. 
Leading logarithms from supersymmetric spectra threshold
corrections to $\alpha_S(\hbox{M}_Z)$ can be summarized in the
following formula \cite{Langacker:1993rq,Carena:1993ag}
\begin{equation}
\Delta \alpha_s^{SUSY}= -\frac{19\alpha_s^2}{28\pi}
\ln\left(\frac{T_{SUSY}}{\hbox{M}_t}\right)
\label{threshSUSY}
\end{equation}
where $T_{SUSY}$ is an effective mass scale given by
\begin{equation}
T_{SUSY}=m_{\widetilde H}
\left({{m_{\widetilde W}}\over{m_{\tilde g}}}\right)^{28\over 19}
\left[\left({{m_{\tilde l}}\over{m_{\tilde q}}}\right)^{3\over 19}
\left({{m_H}\over{m_{\widetilde H}}}\right)^{3\over 19}
\left({{m_{\widetilde W}}\over{m_{\widetilde H}}}\right)^{4\over 19}
\right]\,.
\label{TSUSY}
\end{equation}
This scale is not simply an average of SUSY masses since it can be
smaller than all the masses of the supersymmetric particles
\cite{Carena:1993ag,Langacker:1993rq}. Large values of $T_{SUSY}$ are 
experimentally preferred because in general they contribute negatively
to $\Delta \alpha_s^{SUSY}$, bringing $\alpha_s(M_Z)$ closer to the
experimental average by an estimated  $|\Delta \alpha_s^{SUSY}|
\leq 0.003$ \cite{Langacker:1995fk}. There is in addition, a finite
contribution from supersymmetric threshold corrections which may be
important if the supersymmetric spectrum is light
\cite{lightthresholds}.  Moreover there is also a small conversion
factor from $\overline{\hbox{MS}}$ to $\overline{\hbox{DR}}$
\cite{DR}, as well as possible contributions coming from non
renormalizable operators which can be induced from physics between the
Planck to the GUT-unification scale \cite{NRO}.

Let us now turn to the important issue of the two loop Yukawa
contribution to the gauge coupling constants RGE. This contribution is
not included in \eq{alfaiApp} and is crucial for our purposes,
providing a correction which is negative and can be important if $h_t$
or $h_b$ are large ($t_{\beta}\approx 1$ or $t_{\beta}\approx 50$
respectively). Making a one-step integration we obtain the approximate
expression
\begin{equation}
\Delta\alpha_s^{YUK} \approx -\frac{\alpha_s^2}{32\pi^3}
\ln\left(\frac{\hbox{M}_{U}}{\hbox{M}_t}\right)
\left\{b_{3t}^{\prime}h^2_t + b_{3b}^{\prime}h^2_b\right\}
\label{DYuktoalfas}
\end{equation}
In the small $\tan\beta$ region, the bottom Yukawa coupling is
negligible compared to the top Yukawa, then we get
$\Delta\alpha_s^{YUK} \approx -0.1 \alpha_s^2h_t^2$, giving us an
estimate of the magnitude of this correction. Note that this
correction is not bigger in the high $\tan\beta$ scenario, where both
Yukawas are large, since they are not as large as the top Yukawa in
the low $\tan\beta$ case.

In contrast, in the $\Slash{R}$--MSSM model, the bottom Yukawa
coupling can be non-negligible for any value of $\tan \beta$
\cite{epsyuk}.  As a result we cannot neglect the bottom quark Yukawa 
coupling, since it can be as large as the top quark Yukawa, especially
if the R--parity violating parameters are large.

\section{The $\Slash{R}$-MSUGRA  Model}

In order to illustrate the essential features of the model, it is
enough to consider a one generation $\Slash{R}$--MSSM
\cite{epsrp:1997,NillHempJoshi,Bisset:1998bt,Nowak:1996dx}, since it 
contains the main ingredients relevant for our present discussion. The
superpotential $W$ is
\begin{equation}
W=W_{MSSM}+W_{\Slash{R}} \ ,
\label{superpot}
\end{equation}
where $W_{MSSM}$ is the familiar superpotential of the MSSM
\begin{equation}
W_{MSSM} = \left[ h_t\widehat{Q}_3  \widehat{H}_u\widehat{U}%
_3+\lambda _{0}^D\widehat{L}_0  \widehat{Q}_3\widehat{D}_3 
+h_{\tau} \widehat{L}_0 \widehat{L}_3\widehat{R}_3 
-\mu_{0}\widehat{L}_0 \widehat{H}_u \right]\,.
\label{supWMSSM}
\end{equation} 
Here we are using the notation $\widehat{L}_{0}\equiv\widehat{H}_d$,
$\mu_0\equiv\mu$, and $\lambda_0^D\equiv h_b$ in the superpotential,
and $v_0\equiv v_d$ for the $\widehat{H}_d$ vacuum expectation
value. This notation is justified because $\widehat{H}_d$ and
$\widehat{L}_3$ have the same quantum numbers. The piece of the
superpotential which breaks R--parity is given by
\begin{equation}
W_{\Slash{R}}=-\mu_{3}\widehat{L}_3 \widehat{H}_u 
\label{RpBreaW}
\end{equation}
where $\mu_3$ is the bi-linear R--Parity violating term (BRpV), denoted
$\epsilon_3$ in ref.~ \cite{epsrp:1997,BRpVmas}.

Notice that we do not generate a tri-linear R--Parity violating (TRpV)
term in models that arise from spontaneous breaking of R-parity. In
fact, even if explicit tri-linear terms were present, for the simple
one-generation case they can always be rotated away into the bi-linear
term given in~\eq{RpBreaW}. In other words, the most general
one-generation explicit SUGRA R--Parity violation model is
characterized by a single parameter, which may be chosen either as
$\mu_3$, or as $\lambda_3^D$ or the sneutrino vev.  The converse is
not true,  BRpV cannot be rotated away in favour of TRpV.

Although the above presentation would be in some sense the simplest
and sufficient for our purposes, it will be useful for us in what
follows to keep a redundant parametrization in which the bi-linear and
tri-linear R parity violating terms coexist.

The scalar potential contains the following relevant soft terms
\begin{equation}
V_{soft}=\left(\begin{array}{c}
               \widetilde{L}_0 \\ \widetilde{L}_3
               \end{array} 
        \right)^{\dagger} 
\left(\begin{array}{cc} 
       M_{L_{0}}^2 & M_{L_{03}}^2 \\ M_{L_{30}}^2 & M_{L_{3}}^2  
      \end{array} 
\right) 
\left(\begin{array}{c} 
      \widetilde{L}_0 \\ \widetilde{L}_3  
      \end{array} 
\right) - 
\left( \mu_{\alpha}B_{\alpha} \widetilde{L}_{\alpha} H_u
      - A^D_{\alpha} \lambda _{\alpha}^D \widetilde{L}_3
      \widetilde{Q}_3 \widetilde{D}_3 + h.c. 
\right)  
\end{equation}
where $M_{L_i}^2$ are the soft mass terms and mixing for the down type
Higgs and slepton fields, $B_{\alpha}$ $\alpha=0,3$, are the bi-linear
soft mass parameters ($B_0$ corresponds to the usual $B$ term in the
MSSM), while $A_{\alpha}^D$ are the tri-linear soft mass parameters
($A_0^D$ is the usual $A_D$ term in the MSSM).
 
The equality of the quantum numbers of the down-type Higgs and tau
lepton $SU(2) \otimes U(1)$ superfields opens the possibility to work
in different basis
\cite{Bisset:1998bt,Nardi:1997iy,Ferran:1998ii,rptalks}. This field
redefinition is defined by
\begin{equation}
{{\widehat L'_0}\choose{\widehat L'_3}}=\left(
\matrix{\cos\alpha_L  & \sin\alpha_L \cr
        -\sin\alpha_L & \cos\alpha_L}\right)
{{\widehat L_0}\choose{\widehat L_3}}
\label{rotation}
\end{equation}
where $\alpha_L$ is the angle of rotation, which in turn induces a
rotation of the $\mu $ --terms.  Under this change of basis the
Lagrangian parameters transform and it is impossible to eliminate
completely the effects of the bi-linear
terms~\cite{epsrp:1997,BRpVmas,rptalks}. Note that
different basis may be convenient for different applications
\cite{Ferran:1998ii}.  

Here we are specially interested to express R--parity violating
effects through basis independent parameters
\bea
v_d=\sqrt{v_0^2+v_3^2}\\
\mu =\sqrt{\mu_0^2+\mu _3^2}\\
\lambda ^D=\sqrt{(\lambda _0^D)^2+(\lambda _3^D)^2}
\eea
From the above we can deduce that the natural generalization of the
MSSM definition of $\tan \beta$ is given by 
\beq
\tan \beta = \frac{v_u}{v_d}, 
\eeq
which is also a basis invariant.  This definition differs from the one
used in ref. \cite{epsrp:1997,epsyuk}, namely $\tan\beta=v_u/v_0$.
There are other invariants which turn out to be very
useful~\cite{Davidson} and are defined as
\bea
\cos \zeta =\frac{\mu _\alpha v_\alpha }{\mu v_d} \\
\cos \gamma =\frac{\lambda _\alpha ^D\mu _\alpha }{\lambda^D\mu }\\
\cos \chi =\frac{\lambda _\alpha ^Dv_\alpha }{\lambda ^Dv_d}
\eea
Note that these three parameters are not independent due to the
trigonometric relation
\begin{equation}
\cos \chi = \cos \left( \gamma - \zeta \right)
\label{eq:trigo}
\end{equation}
The remaining R--parity violating variables $\sin \zeta $ and $\sin
\gamma $ determine the $\nu _\tau $ mass and the R--parity violating 
effects in general in the fermionic sector, while $\sin \chi $
characterizes the R--parity violating effects on $\alpha_s$.  As we
will see below there is only one of these parameters which survives,
owing to the minimization conditions of the theory.

In this model the top and bottom quark masses are given by
\begin{equation} 
\label{eq:topmas}\hbox{M}_t=\frac{h_t}{\sqrt{2}}v_u=
 s_{\beta}h_t \frac{\sqrt{2}\hbox{M}_W}{g} 
\end{equation} 
\begin{equation} 
\label{eq:botmas}\hbox{M}_b=\frac 1{\sqrt{2}}\left( \lambda _0^Dv_0+\lambda 
_3^Dv_3\right) =
c_{\beta}c_\chi\lambda^D\frac{\sqrt{2}\hbox{M}_W}{g}
\end{equation} 
This formula for the bottom quark mass is specially interesting, since
it is expressed in terms of basis-independent R--parity violating
effects  parameters.
 
As in the MSSM case to connect the phenomenology at the electroweak
scale with the SUGRA parameter space we need to use the
renormalization group equations.  A question immediately arises as to
the number of additional parameters necessary to characterize the
model. For a one-generation model with universality of soft parameters
at the unification scale only one additional parameter is needed in
addition to the MSUGRA parameters ~\cite{epsrp:1997}.  We have,
however, some freedom in this choice.  To compute the Lagrangian
parameters at the electroweak scale we can follow two different
approaches
\cite{Ferran:1998ii}
\begin{itemize}
\item
the bi-linear or $\mu_3$--approach, in which the parameters which
fix the model are: $$
\left( \hbox{A}_0,\hbox{M}_0,\hbox{M}_{1/2},\hbox{t}_\beta ,\mu _3^U\right)   
$$ Because of the form of the RGE for $\lambda^D_3$, $d\lambda^D_3/dt
\propto \lambda^D_3$, if $\lambda^D_3$ is zero at the unification scale 
it will be zero at the electroweak scale
\item
The second possibility is the $\lambda^D_3$--approach, in this case
the fundamental parameters of the model are $$
\left( \hbox{A}_0,\hbox{M}_0,\hbox{M}_{1/2},\hbox{t}_\beta ,
(\lambda _3^D)^U\right).$$ In contrast to the previous case here one
arrives at the electroweak scale to the coexistence of bi-linear and
tri-linear R--parity breaking parameters.
\end{itemize}
It does not matter which approach we follow because both are
equivalent. Notice that, while in the bi-linear approach one can ignore
tri-linears without loss of generality, the converse is not true: one
can not neglect bi-linears consistently due to the structure of the
RGES. One can change from one basis to another and thus compare
calculations which have been performed in different basis. These
results have to be the same.

Now we are ready to understand how R--parity violation can affect the
gauge coupling unification through the two loop Yukawa contribution to
the RGES for $\alpha_s$. One finds,
\beq
\Delta\alpha_s^{YUK} \approx -\frac{\alpha_s^2}{32\pi^3}
\ln\left(\frac{\hbox{M}_U}{\hbox{M}_t}\right)
\left\{b_{3t}^{\prime}h^2_t + b_{3b}^{\prime}(\lambda^D_0)^2+
b_{3b}^{\prime}(\lambda^D_3)^2 \right\}
\eeq
Where one notes the appearance of the R--parity violating coupling
$\lambda^D_3$.  Clearly this term combines with $\lambda_0^D$ to form
the basis invariant $\lambda^D$ as follows,
$$
\Delta\alpha_s^{YUK} \approx -\frac{\alpha_s^2}{32\pi^3}
\ln\left(\frac{\hbox{M}_U}{\hbox{M}_t}\right)
\left\{b_{3t}^{\prime}h^2_t + b_{3b}^{\prime}(\lambda^D)^2
\right\}
$$
Using the formulas (\ref{eq:botmas},\ref{eq:topmas}) for the top and
bottom masses we obtain 
\beq
\Delta\alpha_s^{YUK} \approx -\frac{\alpha_s^2}{32\pi^3}
\ln\left(\frac{M_U}{M_t}\right)
\frac{g^2}{2M^2_W}(1+t_{\beta}^2)
\left\{b_{3t}^{\prime}\frac{\hbox{M}_t}{t_{\beta}} + 
b_{3b}^{\prime}\frac{\hbox{M}_b}{c_{\chi}} \right\}
\label{maineq}
\eeq

We are now set to demonstrate the direct correlation between the last
term in \eq{maineq} and the magnitude of R--parity violation which, as
already mentioned, is characterized by a unique parameter in this
model. To see this we must make use of the three minimization
equations of the scalar potential of the theory, the two of the MSSM
plus a third equation involving the vev for the tau sneutrino. Using
this equation one can find a relation between $\sin\zeta$ and
$\sin2\gamma$ which finally reduces the extra number of parameters to
simply one, when compared with the R-parity conserving supergravity
model. At first order in $\mu_3/\mu$ can be simplified to
\begin{equation}
\sin\zeta = 
\frac{\mu _0\mu _3}{\mu^2} \left( \delta_B t_{\beta} \pm \delta_M 
\right)= \frac{1}{2}\sin(2\gamma) \left( \delta_B t_{\beta} \pm \delta_M 
\right)  
\label{eq:szeta}
\end{equation}
where
$$ 
\delta_B = \frac{\mu \Delta B} {\left( M_{\widetilde{\nu_3 }}^2 -%
\frac{\mu_3^2}{\mu^2}\Delta M^2 \right)} \qquad \delta_M = \frac{\Delta M^2} 
{\left( M_{\widetilde{\nu_3 }}^2 -\frac{\mu_3^2}{\mu^2}\Delta M^2 \right)} 
\ ,  
$$ 
and we have defined
$$
\Delta B= B_3 -B_0 \qquad \Delta M^2 = M^2_3 - M^2_0
$$ 

We notice that the double sign in eq.(\ref{eq:szeta}) is the result of
the solution to a quadratic equation in the minimization conditions of
the scalar potential. In models with universality of soft terms,
$\delta_M$ is positive but $\delta_B$ can take either sign.

Thus \eq{eq:szeta} shows that, as anticipated, only one of the three
parameters $\zeta,\gamma,\chi$ is independent.  Together with the
above SUGRA parameters which fix the model it determines the Majorana
mass for the tau neutrino. The latter is induced by the mixing of the
original tau neutrino field with the neutralinos~\cite{rpos,rpre} and
is given mainly by the parameter $\sin\zeta$ through the approximate
relation
\begin{equation} 
\label{eq:neumas}
\hbox{M}_{\nu _\tau }^{\nu-\chi mixing}=
\frac{\hbox{M}_Z^2M_{\widetilde{\gamma }}\mu s_\zeta ^2c_\beta ^2}
{\left( 
M_Z^2 M_{\widetilde{\gamma }}s_{2\beta }c_\zeta 
-M_1M_2\mu\right)}
\end{equation} 
which depends on the SUGRA parameters, where we have defined the
parameter $M_{\widetilde{\gamma }} \equiv c_W M_1 + s_W M_2$.  From
\eq{eq:trigo}, \eq{eq:szeta} and (\ref{eq:neumas}) it is evident that 
we can get an expression for $\cos\chi$ whose exact form is
unimportant for our present argument, except for the property that 
$$
\cos\chi \to 1 \:\: \mbox{as}   \:\: M_{\nu_\tau} \to 0
$$ 
Thus the maximum value $c_{\chi} = 1$ corresponds to the R-parity
conserving case.  From this it is clear that the larger the R parity
violation the larger will be the additional contribution coming from
the ratio $\hbox{M}_b/c_{\chi}$ in \eq{maineq}.  The above equation
establishes a relationship that the basis-independent parameter
$c_{\chi}$ bears with the tau neutrino mass.

\begin{figure}[ht] 
\centerline{ \epsfxsize 5.8 truein \epsfbox {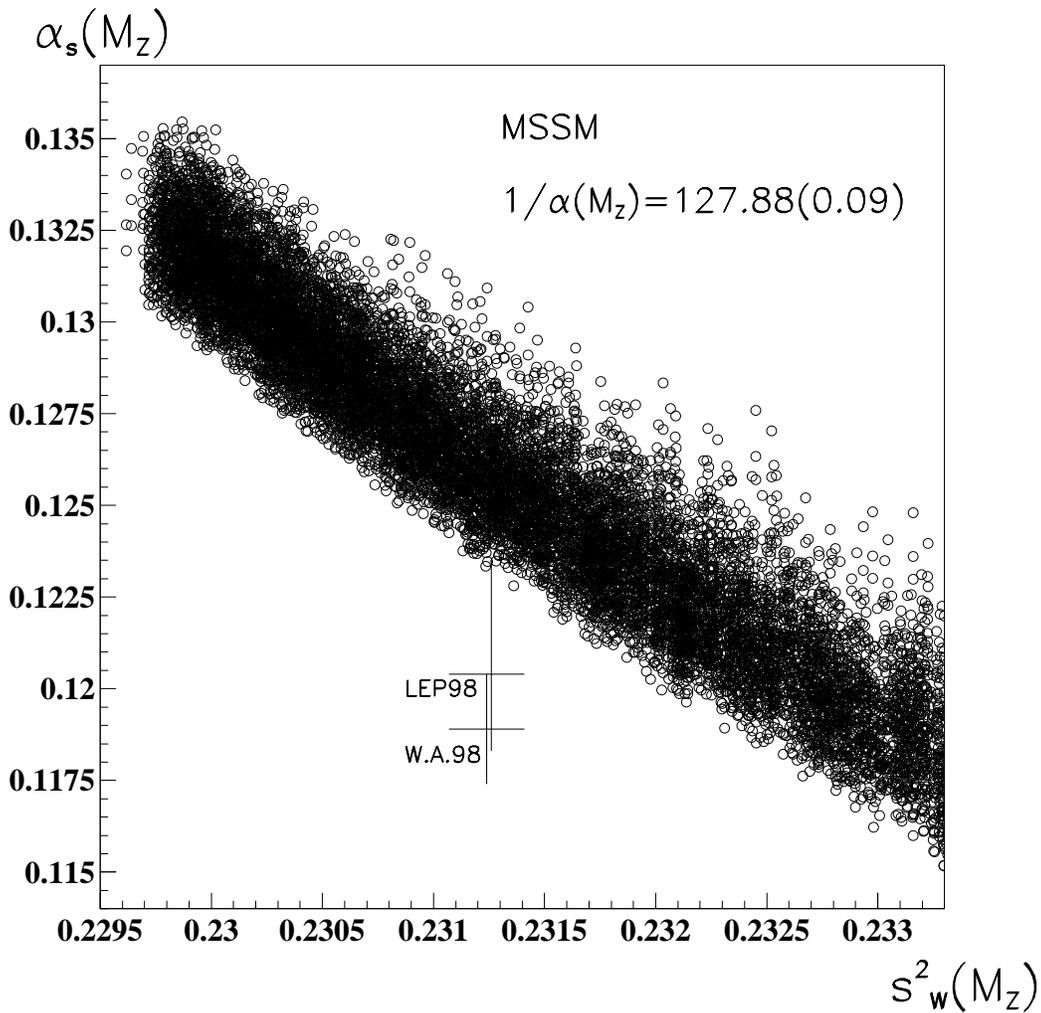}} 
\caption{\it 
$\alpha_s(M_Z)$ versus $\widehat{s}_{Z}$ for the MSSM}  
\label{fig:alfaMSSM}  
\end{figure}  

We now turn to the implications of R-parity violation on the
$\alpha_s$ predictions derived from \eq{maineq} and to our numerical
results.  We have used the two-loop renormalization group equations
for the gauge coupling constants and the Yukawa couplings and the
one-loop RGE for $\mu$--terms and for the rest of the soft parameters
\cite{RMSSMRGE}.  We will study the prediction for the gauge coupling
constants at the $M_Z$ scale in a model with universality of the soft
terms at the unification scale
\footnote{
For the sake of generality and in order to simplify the discussion we
will neglect possible GUT threshold contributions, as well as 
non--renormalizable operator contributions.}.  We compare masses and
couplings at the $M_Z$ scale with their experimental values (see
appendix for a detailed description of the method we have used for
the running of the effective masses to their pole values and the
running of the gauge couplings to their $\overline{\hbox{MS}}$ values
at the  $M_Z$ scale).  

As a first step in our study of the supersymmetric
$\alpha_s(\hbox{M}_Z)$ and $\widehat{s}^2_Z$ predictions we have
updated the standard MSUGRA prediction taking into account the latest
PDG experimental values for $\widehat{\alpha}(\hbox{M}_Z)^{-1}$
\cite{Caso:1998tx} $$
\widehat{\alpha}(\hbox{M}_Z)^{-1}= 127.88\pm0.09
$$
On the other hand for the top, bottom and tau pole masses 
we have used~\cite{Caso:1998tx}
$$
\hbox{M}_t^{pol}=173\pm5.2 \quad\hbox{GeV}
$$
$$
\hbox{M}_b^{pol}=4.1 \quad \hbox{to} \quad 4.4 \quad \hbox{GeV}
$$
$$
\hbox{M}_{\tau}^{pol}=1777.05
\left. 
\begin{array}{c} 
+0.29 \\ 
-0.26
\end{array}\right. \hbox{MeV}
$$ In figure (\ref{fig:alfaMSSM}) we display updated the MSUGRA
prediction for $\alpha_s(\hbox{M}_Z)$ and $\widehat{s}^2_Z$ given as a
scatter plot, where each point corresponds to a different choice of
SUGRA parameters, varying over a wide range, given as
\begin{equation}
\begin{array}{c}
0<\hbox{M}_0< 500 \quad \hbox{GeV} \\
0<\hbox{M}_{1/2}< 500\quad \hbox{GeV} \\
-1000<\hbox{A}_0<1000\quad \hbox{GeV} \\
2 \lsim \hbox{t}_{\beta}<60\qquad\qquad
\end{array}
\label{eq:SUGRA2}
\end{equation}
In the figure one can appreciate the difference between the present
world average for $\alpha_s(\hbox{M}_Z)$ 
$$
\alpha_s(\hbox{M}_Z)^{W.A.}=0.1189\pm 0.0015
$$
and the 1998 average of the LEP measurements ~\cite{Caso:1998tx}
$$
\alpha_s(\hbox{M}_Z)^{LEP98}=0.1214\pm 0.0031
$$
For a discussion on the question of the average of values of
$\alpha_s$ deduced at different energy scales, see references
\cite{Langacker:1993rq,alphaspol}.

We notice that if we fix $\widehat{s}^2_Z$ inside its experimental
range
\footnote{We have moved slightly the $\widehat{s}_Z^2$ for one of the
measurements in order to observe clearly the difference in
the$\widehat{s}_Z^2$ values.},
$$
(\widehat{s}^2_Z)^{W.A.}= 0.23124\pm 0.00024
$$
the MSUGRA $\alpha_s(\hbox{M}_Z)$ prediction lies in the range
$\alpha_s(\hbox{M}_Z) \approx 0.127\pm 0.003$, which is a bit more
than $2\sigma$ higher than the world average.  

\begin{figure}[ht] 
\centerline{ \epsfxsize 5.8 truein \epsfbox {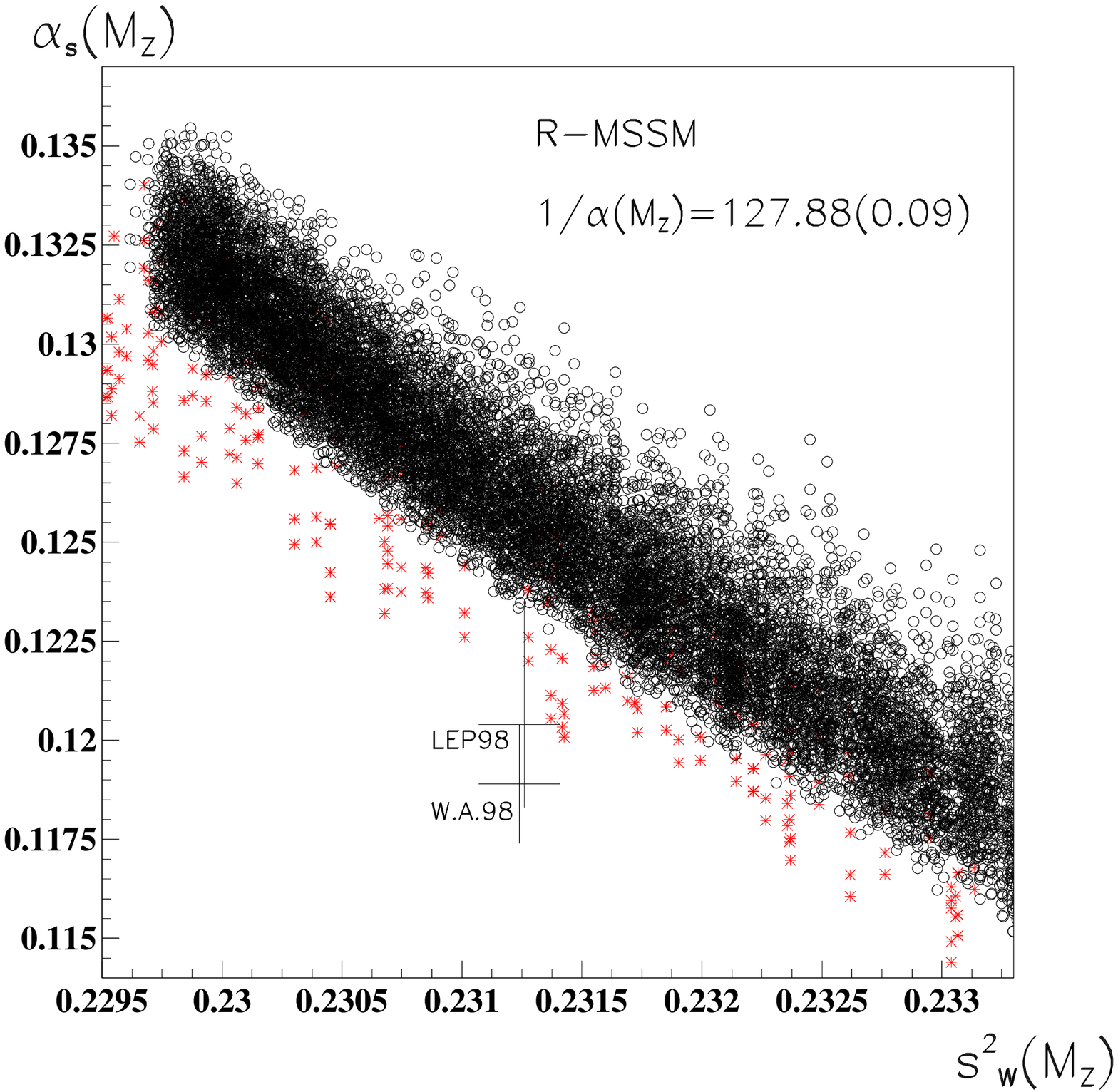}} 
\caption{\it 
$\alpha_s(\hbox{M}_Z)$ versus $\widehat{s}_{Z}$ for the 
$\Slash{R}$--MSSM model}  
\label{fig:alfaRMSSM}  
\end{figure}  

Now we turn to discuss the results we obtain in our bi-linear R-parity
breaking model, $\Slash{R}$--MSSM for short. The method we have used
is similar to the previous procedure.  In this case additional
complications appear because of the mixing between charginos and the
tau lepton and we need to ensure that the tau mass is the
experimentally measured. On the other hand the mixing between the
neutralinos and the neutrino implies the appearance of a mass for the
tau neutrino which arises from the mixing with the neutralinos
~\eq{eq:neumas} and we must ensure that it lies below the experimental
bound~\cite{labbound}.  As we have already seen, the non-zero tau
sneutrino vev implies that we have to take into account the additional
constraint given by the third minimization equation.  Once we satisfy
all these constraints we find that the $\Slash{R}$-MSUGRA predicts
$\alpha_s(\hbox{M}_Z)$ values nearer the experimental value than the
R-parity conserving MSUGRA case.  This comes from the enhanced
negative two-loop bottom-quark Yukawa contribution to the RGE's. For
example, taking the world average experimental value of
$\widetilde{s}_Z^2$, one can move $\alpha_s(\hbox{M}_Z)$ from a
minimum value of approximately $0.125$ in the MSUGRA case down to a
minimum value of $0.122$ or so in the $\Slash{R}$--MSSM model,
bringing closer to the W.A. and within one $\sigma$ from the most
recent average of LEP measurements given in ref. ~\cite{Caso:1998tx}.
These results can be clearly seen from figure (\ref{fig:alfaRMSSM}),
where each point represents a different parameter choice in the
$\Slash{R}$--MSUGRA model.  Notice that the $\Slash{R}$--MSUGRA model
is totally fixed if we know the $\nu_{\tau}$ mass. We have varied the
tau neutrino mass below the laboratory bound $\hbox{M}_{\nu_{\tau}} <
18.2$ MeV \cite{labbound}.

\section{Discussion: $\Delta\alpha_s$ versus \mnt } 

As we have seen, the effect on $\alpha_s(\hbox{M}_Z)$ in the
$\Slash{R}$--MSUGRA model is is related with the $\nu_{\tau}$ mass.
What is the price to decrease $\alpha_s(\hbox{M}_Z)$? Clearly points
with large values for the R--parity violating parameter $\mu_3$ will
have small $\alpha_s(\hbox{M}_Z)$. Typically these points also have
large mas for the tau neutrino. In fact all the points with low
$\alpha_s(\hbox{M}_Z)$ in fig.2 correspond to large \nt masses close
to the present laboratory bound. For these masses many R--parity
violating phenomena have large rates.  Among the latter is the decay
of the lightest neutralino, which will typically occur inside the LEP
and LHC detectors, for reasonable values of parameters.

Large tau neutrino masses would appear in conflict with the
\nm to \nt oscillation interpretation of the recent atmospheric 
neutrino data from underground detectors \cite{Fukuda:1998mi}.  First
we point out that, as it stands, the present data allow for alternative
explanations either in terms of flavour-changing neutrino matter
interactions~\cite{Gonzalez-Garcia:1998hj} or \nm deutrino decay to
\nt~\cite{Barger:1998xk} or \nm decau to \ne \cite{valencia:decayatm}
which might be relevant in the present model or in the presence of a
majoron.  Barring an enhanced statistics up-going muon event sample,
it is hard to dismiss such alternative explanations of the atmospheric
data on the basis of the present information on rates and zenith angle
distributions both of sub as well as multi-GeV events.

Can our result on $\alpha_s(\hbox{M}_Z)$ survive with \nt masses in
the range indicated by the standard oscillation interpretation of the
Superkamiokande results?  In the one-generation approximation the \nm
and the \ne are massless, but it is known that they acquire non-zero
masses due to loops. Barring the results of a detailed investigation
of the loop-generated masses of the two low-lying neutrinos, one can
not give a precise and complete answer to this question. However we
can state that, neglecting these masses one can have a sizeable drop
in $\alpha_s(\hbox{M}_Z)$ for \nt mass in the range close to $3 \times
10^{-2}$ eV, indicated by the best fit of the oscillation hypothesis
~\cite{Gonzalez-Garcia:1998vk}. 

To better understand this statement we make a few approximations.
Consider first eq.~(\ref{maineq}). As we mentioned before, in BRpV the
term proportional to $m_t$ and the term proportional to $m_b$ can be
simultaneously large. In this case, with the two terms of similar
magnitude we have
\begin{equation}
\cos\chi \approx \frac{M_b}{M_t}t_{\beta} \approx 0.017 t_{\beta}
\label{chiapprox}
\end{equation}
which is a necessary condition for large Yukawa contributions to
$\alpha_s$ in BRpV. On the other hand, it is convenient to rewrite the
formula for the neutrino mass in eq.~(\ref{eq:neumas}) by introducing
the mass parameter $\Lambda$ defined by the equation
\begin{equation}
\sin \zeta \equiv \frac{1}{c_{\beta}} \sqrt{ \frac{M_{\nu}}{\Lambda} }
\label{zetalambda}
\end{equation}
where the neutrino mass $M_{\nu}$ is in eq.~(\ref{eq:neumas}) and
$\Lambda={\cal O}(M_Z^2/M_{1/2})$. Therefore, for a neutrino mass of
the order of $0.1$ eV we need $\sin\zeta \approx 10^{-5}
\sqrt{\Lambda} /c_{\beta}$ with $\Lambda$ in GeV, indicating that the
parameter $\sin\zeta$ is very small. In this way, from
eq.~(\ref{eq:trigo}) we see that small neutrino mass implies
$\cos\chi\approx\cos\gamma$. Using this last relation in
eq.~(\ref{eq:szeta}) we find a second expression for $\sin\zeta$:
\begin{equation}
\sin\zeta \approx s_{\chi}c_{\chi}(\delta_Bt_{\beta} \pm \delta_M)
\label{zetachi}
\end{equation}
where the $\delta$'s are defined below eq.~(\ref{eq:szeta}). The
quantity in parenthesis is a good measure of the amount of cancelation
necessary in order to have a sizable effect on $\alpha_s$ with small
neutrino mass.  The cancelation can occur with either sign since the
sign of $\delta_B$ is not fixed. We have that:

\begin{equation}
\delta\equiv (\delta_Bt_{\beta} \pm \delta_M)\approx
\frac{1}{s_{\chi}c_{\chi}c_{\beta}}\sqrt{\frac{M_{\nu}}{\Lambda}}\,.
\label{finetunning}
\end{equation}
We note that in SUGRA with universality of soft SUSY breaking
parameters at unification $\delta_B$ is usually smaller than
$\delta_M$. As a result, the cancellation necessary in order to obtain
small neutrino mass favours large $\tan\beta$ values.  For example,
for $\tan\beta=40$, $c_{\chi}\approx s_{\chi}\approx 0.7$, and a 0.1
eV neutrino mass we have that for $M_{1/2}=200$ GeV the amount of
cancelation is $\delta\approx 1\times 10^{-4}$. If the gaugino mass
parameter is increased to $M_{1/2}=1000$ GeV, the cancellation is
$\delta\approx 3\times 10^{-4}$.  Our approximation is conservative
since we have assumed $\delta_M$ of order 1. However, $\delta_M$ can
be smaller because it is zero at the unification scale and arises only
from the RGE evolution from unification to the weak scale, typically,
$\delta_M \lsim \delta_M \lsim \mbox{few}$ \%. We do not think that
this is a fine tuning. In fact we remind the reader that a similar
amount of cancelation between vev's is already present in the MSSM at
high values of $\tan\beta$.

In short, while our main result on $\alpha_s(\hbox{M}_Z)$ does favour
large \nt masses, there is a range of parameters, motivated by
universality of the soft breaking terms, where the effect naturally
survives even if the \nt mass is rather low.  This guarantees also
that the lightest neutralino would typically decay inside the
detectors now under discussion, changing completely the phenomenology
of supersymmetry from that expected in the MSSM.

\section{Conclusion} 

In conclusion, we have shown how minimal R--parity violating
supergravity can lower the $\alpha_s(\hbox{M}_Z)$ prediction with
respect to the case with conserved the R--parity, as suggested by the
present experimental world average.  We have identified the source of
this effect on the $\alpha_s$ prediction as coming from the two-loop
bottom Yukawa coupling contribution to the renormalization group
evolution of the gauge couplings. This contribution can not be
neglected if the R--parity violating parameters are sizeable. We have
also shown how this effect on the $\alpha_s$ prediction is in general
directly correlated to the value of the tau neutrino mass which is
generated by the mixing of neutralinos and neutrinos.  We have also
discussed to which extent this correlation depends on the initial
conditions for the soft supersymmetry breaking parameters at the
unification scale. We showed how to obtain a sizeable effect on
$\alpha_s(\hbox{M}_Z)$ even in the case that the \nt mass lies in the
range indicated by the simplest neutrino oscillation interpretation of
the atmospheric neutrino data.

\section*{Acknowledgements} 

This work was supported by DGICYT under grants PB95-1077 and by the
TMR network grant ERBFMRXCT960090 of the European Union. M. A. D. was
supported by a DGICYT postdoctoral grant and partly by the
U. S. Department of Energy under contract number DE-FG02-97ER41022.
J. F. was supported by a Spanish MEC FPI fellowship.

\appendix
\renewcommand{\theequation}{\thesection.\arabic{equation}}
\def\sectionapp#1{\setcounter{equation}{0}
\section{#1}}

\section{Appendix: Numerical procedure}

In this appendix we describe with some detail the method we follow to
predict the gauge coupling constants at the $\hbox{M}_Z$ scale.  We
have used the 2--loop RGE's for the gauge coupling constants and for
the Yukawa couplings including R--parity violating couplings
\cite{RMSSMRGE}. We neglect the Yukawa couplings of the first two 
generations.  For the rest of the parameters of the $\Slash{R}$--MSSM
model we have used 1--loop RGE's \cite{RMSSMRGE}.  We have imposed
universality of soft parameters and gauge coupling unification at a
scale $\hbox{M}_U$.  We have explored two different values for the
unification scale, $\hbox{M}_U$ ($1.2 \times 10^{16} < \hbox{M}_U <
3.6 \times 10^{16} \hbox{GeV}$), and for the gauge coupling constant
at the unification scale $\alpha_U$, ($ 23.5 < \alpha_U^{-1} < 24.5$).
Using the RGE's we have found the gauge coupling constants at
$\hbox{M}_t$ and then we have evolved down to $\hbox{M}_Z$ scale as
explain below. On the other hand we have computed the pole masses from
the running masses at $\hbox{M}_t$ following the same procedure as the
ref. \cite{Barger:1993ac}.  First of all we have to explain how we
compute the Yukawa couplings at $\hbox{M}_t$ at the SM side.  We have
to use the right matching conditions at $\hbox{M}_t$ which are easy to
compute from the formulas (\ref{eq:topmas}) and (\ref{eq:botmas}) for
the $h_t$, $\lambda^D$ y $h_{\tau}$ Yukawas. In the $\Slash{R}$--MSSM
model are \cite{epsyuk}

$$
h_t(\hbox{M}_t)^{SM}= s_{\beta}h_t(\hbox{M}_t)^{\Slash{R}}
$$
$$
h_b(\hbox{M}_t)^{SM}=c_{\chi}c_{\beta} \lambda^{D}(\hbox{M}_t)
$$
$$
h_{\tau}(\hbox{M}_t)^{SM}=\frac{c_{\beta}}
{\left(1-s^2_{\zeta}f(M_2,t_{\beta},\mu,c_{\zeta})\right)^{1/2}}
h_{\tau}(\hbox{M}_t)^{\Slash{R}}
$$

These conditions reduce to the MSSM matching conditions in the limit
$c_{\zeta}$,$c_{\chi}\rightarrow 1$.

In order to run of masses and couplings to their experimental values
we use known relations.  First we have evolved $\alpha_1$ and
$\alpha_2$ from scale $\hbox{M}_t$ to scale $\hbox{M}_Z$ to compute
$\alpha(\hbox{M}_Z)$ and $\widehat{s}^2_Z$.  For $\alpha_s$, given the
value $\alpha_s(\hbox{M}_t)$, which we get from the running of the
RGE's from the unification to the $\hbox{M}_t$ scale, we can compute
$\Lambda_{QCD}$ at $\hbox{M}_t$ using the approximate solution for
$\alpha_s$ in the SM \cite{alf3loop} which includes 3--loop QCD
contributions $$
\alpha _s(\mu )=\frac \pi {\beta _0 t }
\left[ 1-\frac{\beta _1}{\beta _0^2}\frac{\ln \left( t \right) }
{ t }+\frac{\beta _1^2}{\beta _0^4 t^2 }\left( \left( \ln \left(t \right) 
-\frac 12\right) ^2+\frac{\beta _2\beta _0}{\beta _1^2}%
-\frac 54\right) \right],   
$$ 
where
$$
t=\ln 
\left(  
\frac{\mu ^2}{\Lambda ^2}
\right)
$$
$$ 
\begin{array}{c} 
\beta _0=\left( 11-\frac 23n_f\right) \frac 14 \\  
\beta _1=\left( 51- 
\frac{19}3n_f\right) \frac 18 \\ \beta _2=\left( 2857-\frac{5033}9n_f-\frac{%
325}{27}n_f^2\right) \frac 1{128} 
\end{array} 
$$ 
Later using the same formula we can extrapolate $\alpha_s$ at $\hbox{M}_Z$.
To compute the top quark pole mass we use \cite{Tarrach:1981}
$$
\hbox{M}_t^{pol}=\hbox{M}_t(\hbox{M}_t)\left[1+
\frac{4}{3\pi}\alpha_3(\hbox{M}_t) \right]
$$
On the other hand to compute the bottom quark pole mass we use the
quark effective mass formula which includes 1--loop QED and 3--loop
QCD contributions
$$ 
\hbox{M}_b(\hbox{M}_t)=\hbox{M}_b(\hbox{M}_b)
\left( \frac{\alpha \left( \hbox{M}_t\right) }
{\alpha \left( \hbox{M}_b(\hbox{M}_b)\right) }\right) 
^{\frac{\gamma _0^{QED}}{b_0^{QED}}}
\frac{F(\alpha_3(\hbox{M}_t))}{F(\alpha _3(\hbox{M}_b(\hbox{M}_b)))},  
$$ 
where the QED beta function and the anomalous dimension, 
$\gamma _0^{QED}$ and $b_0^{QED}$, are given by\cite{Hall:1980}
$$ 
\gamma _0^{QED}=-3Q_f^2  
$$ 
$$ 
b_0^{QED}=\frac 43\left( 3\sum Q_u^2+3\sum Q_d^2+\sum Q_e^2\right)   
$$  
where the sum runs over all the active fermions at the relevant scale.
The formula $F\left(\alpha_s\left(\mu\right)\right)$ is given by
\cite{alf3loop}
$$ 
F\left(\alpha_s\left(\mu\right)\right)=
\left( \frac{2\beta _0\alpha _s\left( 
\mu \right) }\pi \right)^{\gamma _0/\beta _0}
\left\{ 1+\left( \frac{\gamma 
_1}{\beta _0}-\frac{\gamma _0\beta _1}{\beta _0^2}\right) \frac{\alpha 
_s\left( \mu \right) }\pi +\frac 12\left[ \left( \frac{\gamma _1}{\beta _0}-%
\frac{\gamma _0\beta _1}{\beta _0^2}\right) ^2+\right. \right.   
$$ 
$$ 
\left. \left. \left( \frac{\gamma _2}{\beta _0}+\frac{\gamma _0\beta _1^2}{%
\beta _0^3}-\frac{\beta _1\gamma _1+\beta _2\gamma _0}{\beta _0^2}\right) 
\right] \left( \frac{\alpha _s\left( \mu \right) }\pi \right) ^2+O\left( 
\alpha _s^3 \left( \mu \right) \right) \right\}   
$$ 
where
$$ 
\begin{array}{c} 
\gamma _0=1 \\  
\gamma _1=\left(  
\frac{202}3-\frac{20}9n_f\right) \frac 1{16} \\ \gamma _2=\left( 1249-\left(  
\frac{2216}{27}+\frac{160}3\zeta (3)\right) n_f-\frac{140}{81}n_f^2\right) 
\frac 1{64} 
\end{array} 
$$ 
Finally to compute tau lepton pole mass from the tau running mass
at $\hbox{M}_t$ we use
$$ 
\hbox{m}_{\tau}^{\hbox{pol}}= \hbox{m}_{\tau}\left( \mu \right)
\left[ 1+\frac{\alpha \left( \mu \right) }%
\pi \left( 1+\frac 34\ln \left( \frac{\mu ^2}{\hbox{m}_{\tau}^2\left( \mu \right) }%
\right) \right) \right]  
$$   

In summary, starting with the basic parameters $\hbox{M}_0$,
$\hbox{A}_0$, $\hbox{M}_{1/2}$, $\hbox{t}_{\beta}$, $\mu_3$,
$\hbox{M}_U$ and $\alpha_{G}$ we have required that
$\alpha(\hbox{M}_Z)$ as well as the top, bottom and tau pole masses
$\tau$ were inside their experimental measurements in order to obtain
a prediction for the variables $\widehat{s}^2_{Z}$ and
$\alpha_s(\hbox{M}_Z)$ which can be seen in figures.

\end{document}